\documentclass[useAMS,usenatbib]{mn2e}
\usepackage{graphicx}
\usepackage{subfigure}
\usepackage{natbib}

\title[Standard model]{The standard model of star formation applied to massive stars: 
accretion disks and envelopes in molecular lines}
\author[E. Keto and Q. Zhang]{Eric Keto$^{1}$\thanks{E-mail:
keto@cfa.harvard.edu (EK); \hfill\break
qzhang@cfa.harvard.edu (QZ)} and Qizhou
Zhang$^{1}$ 
\\
$^{1}$Harvard-Smithsonian Center for Astrophysics, 160 Garden St, Cambridge, MA 02420, USA 
}
\begin{document}

\date{May 8, 2009}


\maketitle

\label{firstpage}

\begin{abstract}
We address the question of whether the formation of high-mass
stars is similar to or differs from that of solar-mass stars through
new molecular line observations and modeling of the 
accretion flow
around the massive protostar IRAS20126+4104. We combine new 
observations of
NH$_3$ (1,1) and (2,2) made
at the Very Large Array, new observations of CH$_3$CN(13-12)
made at the Submillimeter Array, 
previous VLA observations of NH$_3$(3,3), NH$_3$(4,4),
and previous Plateau de Bure observations of
C$^{34}$S(2-1), C$^{34}$S(5-4),
and CH$_3$CN(12-11) 
to obtain a data set of molecular lines covering 15 to 419 K in
excitation energy. We compare these observations
against simulated molecular line spectra predicted from a model
for high-mass star formation based on a scaled-up version of the
standard disk-envelope paradigm
developed for accretion flows around low-mass stars.
We find that in accord with the standard paradigm, the observations require
both a warm, dense, rapidly-rotating disk and 
a cold, diffuse infalling envelope. This study suggests that 
accretion processes around 10 M$_\odot$ stars
are similar to those of solar mass stars.
\end{abstract}

\begin{keywords}
Keywords.
\end{keywords}

\section{Introduction}

Does the formation of massive stars differ 
significantly from
that of solar mass stars?  
As far as we know,
stars of all masses form in gravitationally unstable regions of
molecular clouds and gain their mass by accretion.
A standard model  developed
for accretion flows around low-mass stars
consists of  two-components, a
rotationally-supported disk inside a freely-falling envelope 
\citep{Shu1987, Hartmann2001}. 
This model has been particularly successful in
explaining infrared observations of low-mass star
formation.
The disk and envelope produce
an excess of long-wavelength infrared emission
that has been adopted as the identifying signature of
accreting protostars in Galactic \citep{Whitney2003}
and extragalactic 
\citep{Whitney2008} star-forming regions.
Furthermore, because the disk and envelope have different
densities and temperatures, the evolutionary state of the protostars
can be identified by the shape of the infrared spectral
energy distribution: class 0 (envelope
dominated) and class I (disk dominated)
\citep{Lada1987, Andre1993}. 

Does this standard two component accretion model
developed for low-mass stars also describe the accretion
flows around massive stars? There are some doubts.
The more massive stars are
luminous enough to generate radiation pressure
and hot enough to ionize their own accretion flows
such that the outward radiative and thermal pressures
rival the inward pull of the stellar gravity
\citep{LarsonStarrfield1971,Kahn1974,Keto2002,KetoWood2006}.
Do these outward pressures result in
accretion flows that are different around more massive
stars? In this paper we compare new and previous 
molecular line observations of an accretion flow around
one massive star 
against the standard disk-envelope paradigm for
accretion flows around low-mass stars.

The previous molecular line observations
of the massive protostar IRAS20126+4104
\footnote{
IRAS20126 is located in the Cygnus-X region at a 
distance 
of 1.7 kpc (Wilking et al. 1989).
}, 
suggest an accretion disk and bipolar outflow
around a 7 to 15 M$_\odot$ protostar
embedded in a dense molecular envelope 
\citep{Cesaroni1997, 
ZhangHS1998,
Cesaroni1999,Kawamura1999,
ZhangHSC1999,
Shepherd2000,
Cesaroni2005, 
Lebron2006,
Su2007,Qiu2008}. 
Previous infrared
observations of absorption and scattering also
reveal a disk and outflow cavity
immediately around the star  
\citep{Sridharan2005,deBuizer2007}. 

In this study, we assemble a suite of
observations of molecular lines of different excitation
temperature in order to 
compare with molecular spectra predicted from
the disk-envelope model.
Lines of different excitation temperature
are useful because
massive stars heat
the surrounding molecular gas to 
observationally significant temperatures ($> 100 K$)
at observationally significant distances from the star,  
($T_{gas}  \sim T_\ast (R/R_\ast)^\alpha$),
and we can exploit the 
relationship between temperature and radius
to distinguish
emission from gas at different radii around the star.
We expect to
identify the
emission from the higher excitation temperature
lines  with gas in the flow closer to the star and also to
separate the emission from the disk and envelope
components.
In previous observations,
\citet{Keto1987} and \citet{Cesaroni1994} used 
this technique, observing
several lines of NH$_3$ 
to study the accretion flows around
very high mass stars associated with HII regions.

We present new 
observations of the NH$_3$ (1,1) and (2,2), 
inversion transitions made with the National Radio
Astronomy Observatory's Very Large Array 
(VLA)\footnote{The National Radio Astronomy
Observatory is a facility of the National Science Foundation
operated under cooperative agreement by Associated Universities,
Inc.}
and new observations
of CH$_3$CN(13-12) made with the Submillimeter Array (SMA). 
The new observations of NH$_3$(1,1) and NH$_3$(2,2)
have a factor of 3 better sensitivity than
the earlier observations of \citet{ZhangHS1998}.
Combined with previous observations
of NH$_3$(3,3) and NH$_3$(4,4) \citep{ZhangHSC1999} the
4 NH$_3$ lines span a range of
excitation temperatures from 23 K to 200 K.
We also have additional lines of lower and higher
excitation temperature from
previous observations of the 
C$^{34}$S(2-1) and C$^{34}$S(5-4) lines
with energies of 7 K and 35 K respectively,
and the ladder of CH$_3$CN (J=12-11) lines with 
energies ranging from
69 K for (K=0) to 419 K for (K=7).
The C$^{34}$S and CH$_3$CN observations were previously
presented in \citet{Cesaroni1999} and \citet{Cesaroni2005}.

We specify the two-component accretion model in terms of
6 parameters: the scale factors for the 
density and temperature of the 
envelope and of the disk, the angular momentum of the
envelope, and the stellar mass. 
We use our molecular line emission code MOLLIE
to predict molecular line spectra from
the parameterized model, and we use
a least-squares fitting procedure
to adjust the parameters
to fit the observations. 

We find that the two-component, disk-envelope model can 
successfully describe
the observations, but a single component model
cannot. A warm, dense rotationally-supported
disk is required to obtain sufficient brightness and width 
in the high-excitation lines, and a cold, large-scale
envelope is required to match the emission from the
lower excitation lines.
We find no evidence that the accretion flow
around IRAS20126
is profoundly altered by the outward force of
radiation pressure or by ionization.
At 10 M$_\odot$, the star might simply not be
luminous enough or hot enough for
its radiation pressure or ionization to
significantly affect the accretion process.
Based on this example, 
the accretion flows around 10 M$_\odot$
stars are quite similar to
flows around lower-mass stars.

\section{The Data}\label{data}

\subsection{Ammonia  inversion lines}

We observed the IRAS 20126+4104 region,
R.A. (J2000) = 20:14:26.06,
decl. (J2000) = 41:13:31.50, in the inversion transitions
of NH$_3$ (J,K) = (1,1), (2,2), (3,3) and
(4,4) on 1999 March 27 and 1999 May 29 with the VLA 
in its D configuration.
We used the 2IF correlator mode to sample both the right and left polarizations,
a spectral bandwidth of 3.13 MHz, and a channel 
width of 24 kHz or 0.3 kms$^{-1}$. The primary beam 
of the VLA at the NH$_3$ line frequencies
is about $2'$. 
Quasars 3C48 and 3C286 were used for flux calibration, 3C273 and 3C84
for bandpass calibration, and quasar 2013+370 
for phase calibrations. The NH$_3$(3,3) and NH$_3$(4,4) observations
were previously discussed in \citet{ZhangHSC1999}.

\begin{figure}
\includegraphics[width=5in,angle=270]{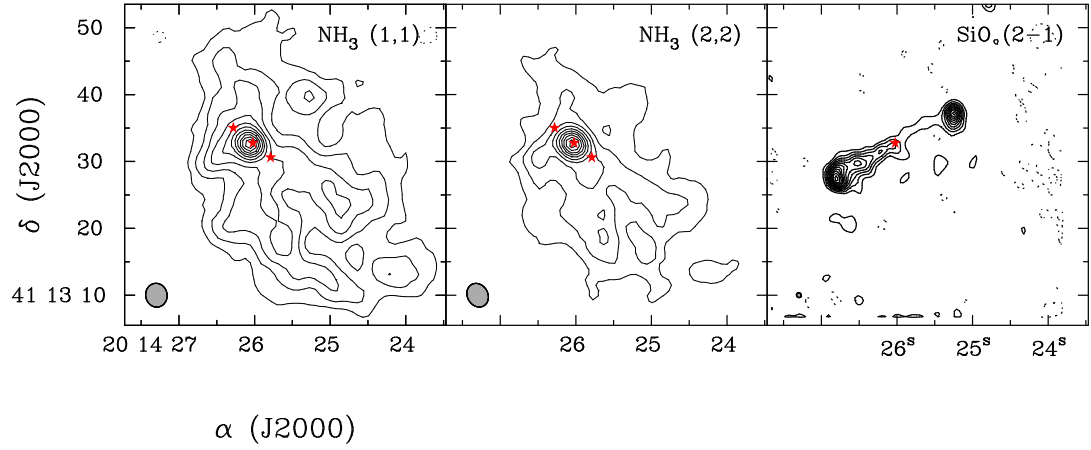}
\vspace{-1in }
 \caption{Images of IRAS20126 in NH$_3$(1,1), NH$_3$(2,2), 
and SiO(2-1). The red stars in the NH$_3$ images show
the locations of the NH$_3$ spectra in figures \ref{fig:nh3disk}
and \ref{fig:nh3nodisk}. See text for absolute coordinates.
The single star in the SiO image
corresponds to the middle star in the other images.
The angular resolution (FWHM) is indicated by the shaded
ellipse, lower-left corner of each panel. The contour interval
and first contour is  35 mJy kms$^{-1}$ for
the NH$_3$(1,1) and NH$_3$(2,2) emission,
and 130 mJy kms$^{-1}$ for the SiO(2-1). 
The SiO image is from \citet{Cesaroni1999}.
}
 \label{fig:maps}
\end{figure}

The VLA data were processed using the NRAO Astronomical Image
Processing System (AIPS) package. Data from the two days were
combined to achieve an rms 
of 3 mJy per 0.3 km~s$^{-1}$ channel for the
NH$_3$ (1,1), (2,2) lines and 
6 mJy and 4 mJy per 0.6 km~s$^{-1}$ channel
for the NH$_3$ (3,3) and (4,4) lines respectively.
Figure \ref{fig:maps} shows two images of IRAS20126 in the integrated intensity
of NH$_3$(1,1) and (2,2). The protostar is located within the bright
core of the larger scale cloud. Also shown in figure \ref{fig:maps} is
the image of the bipolar outflow in SiO(2-1) from \citet{Cesaroni1999}.
The NH$_3$ spectra used in the analysis are taken at positions offset from
the phase center 
by -0.23, 1.00 (center),
-3.10, 3.25 (left),
and 2.50,-0.94 (right)
in arc seconds of RA and dec. 
The corresponding absolute positions are
R.A. (J2000) = 20:14:26.02,
decl. (J2000) = 41:13:32.8 (center) ,
R.A. (J2000) = 20:14:25.79,
decl. (J2000) = 41:13:30.8 (right) ,
and 
R.A. (J2000) = 20:14:26.28,
decl. (J2000) = 41:13:35.0 (left) ,
The locations of the spectra are shown as
red stars on figure \ref{fig:maps}.

\subsection{Methyl Cyanide and Carbon Sulfide lines}

The CH$_3$CN(13-12) observations were made with the SMA on July 10, 2006
with a maximum baseline of
$\sim 0.5$ km, a beam size of $0.40 \times 0.36$",
and a phase center of R.A. (J2000) = 20:14:26.02,
decl. (J2000) = 41:13:32.7. The channel width is 0.8 kms$^{-1}$
at the observing frequency of 239 GHz. The rms noise of the data
is 0.05 Jy beam$^{-1}$ channel$^{-1}$ or 8 K channel$^{-1}$
in brightness temperature. 

Data in the CH$_3$CN J = 12-11, and C$^{34}$S J = 5-4 and J = 2-1 transitions
were obtained from the IRAM interferometer between 1996 December through 1997 March,
and 2002 January through March. The phase center of the observations is
R.A. (J2000) = 20:14:26.03,
decl. (J2000) = 41:13:32.7. The primary beam of the array at the 
frequency of the observed CH$_3$CN transitions (230 GHz) and C$^{34}$S(5-4) (241 GHz) is 
approximately $20^{\prime\prime}$ with a spatial resolution
of about  $0.8^{\prime\prime}$. At the lower frequency of the  C$^{34}$S(2-1) transtion (96 GHz)
the primary beam is about $50^{\prime\prime}$ and the spatial resolution is
and $2.4^{\prime\prime}$.
The full details of these observations can be found
in Cesaroni et al. (1999, 2005).

\section{The Standard Model for Star Forming Accretion Flows}

\subsection{The density of the accretion flow}

The standard model for a star-forming accretion flow consists of
an inflowing envelope around a rotationally supported disk.
Similar to the approach in \citet{Whitney2003},
we use the model of  \citet{Ulrich1976} to describe the gas density  in the
infalling envelope and standard thin-disk theory \citep{Pringle1981} to
describe the disk, 

The model for the envelope assumes that the gas flows toward
a gravitational point source along ballistic trajectories. The flow 
conserves angular
momentum and ignores the pressure and self-gravity of the gas.
This is the same envelope model that \citet{Whitney2003} 
adopted for low-mass stars and similar to the model 
used to describe the accretion flow onto a cluster of high-mass stars, G10.6-0.07
\citep{Keto1990, KetoWood2006}. Thus despite, or perhaps because of,
its simplicity, the model has found application to
accretion flows from low-mass stars to star clusters.  
The model is particularly
useful in analyzing observational data because only
3 parameters are required to describe individual cases:
1) the density of the gas at a single radius (the gas density elsewhere follows
from mass conservation), 2) the mass of the point source, 
and 3) the specific angular momentum.  

The model for the envelope is fully described in \citet{Ulrich1976}. 
We use the equations as presented in \citet{Keto2007}, and \citet{Mendoza2004}.
The envelope density is  (equation 10 of \citet{Keto2007}),
\begin{eqnarray*}
\rho_{\rm env} = \rho_{e0} (r/R_d)^{-3/2}\bigg(1+{{\cos{\theta}}\over{\cos{\theta_0}}}\bigg)^{-1/2}
\end{eqnarray*}
\begin{equation}\label{eq:env_density}
\quad\quad\quad\times \quad (1 + (r/R_d)^{-1}(3\cos^2{\theta_0} - 1))^{-1}
\end{equation}
where $\rho_{e0} = \rho_{\rm env}(R_d)$ is the density in the mid-plane, $\theta=\pi/2$, 
at radius, $R_d$, 
$\theta_0$ is the initial polar angle of the streamline, and $\theta$ and $r$  are the 
polar angle and radius at
each point along a streamline. The angle, $\theta$ is related to the
polar radius, $r$, and the initial angle $\theta_0$ (equation 7 of \citet{Keto2007}),
\begin{equation}
r = { R_d\cos{\theta_0 }  \sin ^2 {\theta_0} \over { \cos{\theta_0} - \cos{\theta}   }}
\label{radius}
\end{equation}
The density is related to the mass accretion rate (equation 11 of \citet{Keto2007})
\begin{equation}\label{eq:accretion_rate}
\dot{M} = \rho_{e0}(R_d) 4\pi R_d^2 v_k
\end{equation}
where $v_k$ is the Keplerian velocity at $R_d$.

The model for the disk
assumes that a rotationally-supported disk forms 
at the radius 
where the centrifugal
force in the rotating 
envelope equals the gravitational force of the 
point mass,
\begin{equation}\label{eq:RD}
{{\Gamma^2}\over{R^3_d}} = {{GM_\ast}\over{R^2_d}}
\label{centrifugal}
\end{equation}
and $\Gamma = v_\phi R$. The disk is truncated at $R_d$.

The gas density in the disk is (equation 3.14 of \citet{Pringle1981}),
\begin{equation}
\rho_{\rm disk}(z,R) =  \rho_{0}(R) \exp(-z^2/2H^2)
\end{equation}
where $\rho_0(R)$ is the density in the mid-plane, $\theta=\pi/2$, at
any radius $R$.
We use $R = \sqrt{x^2 + y^2}$ to denote the cylindrical radius, and 
$r=\sqrt{x^2 + y^2 + z^2}$ for the polar radius.
In the thin disk theory, the scale height, $H$, is
\begin{equation}
H^2 = c_s^2 R^3/GM_\ast
\end{equation}
where $c_s$ is the sound speed. We follow \citet{Whitney2003} and
use a modified version of this
equation,
\begin{equation}
H = H_0 (R/R_\ast)^{1.25}
\end{equation}
with $H_0 = 0.01 R_\ast$. 
The density in the mid-plane of the accretion disk, $\rho_0(R)$, is
\begin{equation}\label{eq:disk}
\rho_0 = \rho_{d0} ( {{R_d} / {R}} )^{2.25}
\end{equation}
where $\rho_{d0}$ is the density in the mid-plane at $R_d$.
Again following \citet{Whitney2003}, we use an exponent of 2.25
in equation \ref{eq:disk} rather than 2 which is derived
in the thin-disk theory. As explained in \citet{Whitney2003}
the modifications to the equations for
the scale height
and mid-plane densities are based
on fits to numerical models of disk structure.

The densities in the disk and envelope at radius $R_d$ are
related by a factor, $A_\rho$,
\begin{equation}\label{eq:disk_factor}
\rho_{d0} = A_{\rho}\rho_{e0}
\end{equation}
For example, in the steady-state flow,
the disk would have a
higher density if, as should be the case, 
the inward velocities in the rotationally-supported
disk were lower than in the freely-falling envelope. Because the
inward velocities in the disk are not described by the thin-disk
theory, we leave
$A_\rho$ as an adjustable parameter.
The total gas density at any point 
in the model is the sum of the densities in the disk and envelope,
\begin{equation}\label{eq:total_density}
\rho_{\rm gas} = \rho_{\rm disk} + \rho_{\rm env}
\end{equation}

\subsection{The temperature}
We assume that the envelope is heated by the star. 
The envelope temperature is (equation 7.36 of \citet{LC1999})
\begin{equation}\label{eq:Tenv}
T_{\rm env} = T_\ast (R_\ast/2r)^{2/(4+p)}
\end{equation}
where $p$ in our model is an adjustable parameter related to the dust opacity and
the geometry of the flow. If the density structure were spherically
symmetric, then $p$ would be the exponent in the frequency dependence, $\nu^p$, of
the Planck mean opacity of the dust.
In a flattened flow, $p$ can be negative if the geometrical
dilution of the radiation
in the accretion flow is greater than $r^{-2}$. 
This can happen 
if the dust at each radius absorbs and
isotropically re-emits the outward flowing radiation. Radiation that is
emitted perpendicular to the disk escapes from the flow. Only the
radiation that is emitted in the direction along the flattened flow
continues to heat the dust at larger radii. The result is 
a decrease in the radiation in the disk faster than
$r^{-2}$.

In the thin-disk theory, the disk is heated by dissipation related 
to the accretion rate.
The disk temperature is (equation 3.23 of \citet{Pringle1981})
\begin{equation}\label{eq:Tdisk}
T_{\rm disk} = B_T \bigg[\bigg({{3GM_\ast\dot{M}}\over{4\pi R^3 \sigma}}\bigg)
	\bigg(1-\sqrt{ {{R_\ast}\over{R}} }\bigg)
	\bigg]^{1/4}
\end{equation}
We include an adjustable factor, $B_T$, to 
allow for additional, or possibly less, disk heating. For example, 
the observations constrain the gas density
through the observed optical depth of NH$_3$. This means that
the density and therefore the accretion rate, $\dot{M}$,  are
dependent on the assumed NH$_3$ abundance which is
not well known. The adjustable
factor, $B_T$, decouples the disk temperature 
from the assumed molecular abundance.
Also the disk temperature may be raised by stellar
radiation (passive heating). The factor, $B_T$, allows
for these effects in an approximate way. 

The gas temperature in the model is the density weighted average of the
envelope and disk temperatures, 
\begin{equation}\label{eq:Tavg}
T_{\rm gas} =  {{ T_{\rm disk} \rho_{\rm disk} + T_{\rm env} \rho_{\rm env} }\over
	 {\rho_{\rm disk} + \rho_{\rm env}  }}  
\end{equation}

\subsection{The velocity}

The 3 components of the gas velocity in the envelope are 
given in spherical coordinates by equations 4,5,6 of \citet{Keto2007},
but there are errors in 
earlier papers. In particular, equations 5 and 6 of \citet{Keto2007}
and equation 8 of \citet{Ulrich1976} are incorrect. Equation 8 of
\citet{Ulrich1976} (same as equation 5 of \citet{Keto2007}) does
not result in energy conservation, $mv^2/2 = GM/r$, when combined
with the other velocity components. The error is small, 1 part in
$10^4$. The following equations, same as in \citet{Mendoza2004},
are exact.
\begin{equation}
v_r(r,\theta) = 
	-\bigg( {{GM_\ast}\over{r}} \bigg)^{1/2} \bigg( 1+ {{\cos{\theta}}\over{\cos{\theta_0}}} \bigg)^{1/2}
\label{eq:vr}
\end{equation}
\begin{equation}
v_\theta(r,\theta) = \bigg( {{GM_\ast}\over{r}} \bigg)^{1/2} 
	\bigg({{\cos{\theta_0} - \cos{\theta}}\over{\sin\theta}}\bigg)
	\bigg( 1+ {{\cos\theta}\over{\cos{\theta_0} }} \bigg)^{1/2}
\label{eq:vtheta}
\end{equation}
\begin{equation}
v_\phi(r,\theta) = \bigg( {{GM_\ast}\over{r}} \bigg)^{1/2} 
	{\sin{\theta_0}\over{\sin{\theta}}}
	\bigg( 1 - {{\cos{\theta}}\over{\cos{\theta_0}}} \bigg)^{1/2}
\label{eq:vphi}
\end{equation}
The velocity in the disk is simply the Keplerian velocity,
\begin{equation}
v_{\rm disk}(R) = \sqrt{GM_\ast/R}
\label{eq:keplerianV}
\end{equation}
where the velocity in the disk is purely azimuthal.
We assume that the radial velocity in the rotationally-supported
disk is comparatively small.
The gas velocity in the model is the density weighted average of the
envelope and disk velocities,
\begin{equation}
\vec{v}_{\rm gas} =  {{ \vec{v}_{\rm disk} \rho_{\rm disk} + \vec{v}_{\rm env} \rho_{\rm env} }\over
	 {\rho_{\rm disk} + \rho_{\rm env}  }} 
\end{equation}
where $\vec{v}_{\rm env}$ is given by equations \ref{eq:vr} through \ref{eq:vphi} and
$\vec{v}_{\rm disk}$ is given by \ref{eq:keplerianV}.

\subsection{The density singularities in the accretion model}\label{singularities}
The density in the Ulrich model 
is singular in the mid-plane of the disk at
the centrifugal radius and also at the origin.
These singularities are caused by the
convergence of the streamlines in the simple mathematical description of the flow.
This convergence is not expected on physical grounds
because gas pressure, neglected in the Ulrich
model, would prevent it.
We handle the singularities in two ways.
We define the computational grid to
have an even number of cells so that the
centers of the middle cells are above and
below the midplane and around the origin. Second, we smooth the
density in the radial direction with
a Gaussian with a width of $R_d/2$.  

\subsection{The model parameters}

Based on the analyses of the previous observations
cited in the introduction, we assume that 
the disk-envelope is viewed edge-on.
The model contains
6 adjustable parameters: \hfill\break
1) $\rho_0$ sets the density of the envelope (equation \ref{eq:env_density}) 
and the mass accretion rate (equation \ref{eq:accretion_rate}), \hfill\break
2) $p$ sets the exponent of the power law decrease of the temperature 
in the envelope (equation \ref{eq:Tenv}), \hfill\break
3) $\Gamma= v_\phi r$ is the specific angular momentum (equation \ref{eq:RD})
of the envelope flow, \hfill\break 
4) $M_\ast$ is the stellar mass \hfill\break
5) $A_\rho$ sets the ratio (equation \ref{eq:disk_factor}) 
of the disk density to the envelope density 
at $R_d$, \hfill\break
6) $B_T$ is factor multiplying the disk temperature (equation \ref{eq:Tdisk}). \hfill\break

\section{Fitting the model to the NH$_3$ data}

The disk enters into the model additively, and we can test 
for the presence of a disk by fitting models to the data 
with and without the disk. In the first case, with the disk, we
adjust all 6 model parameters for both the envelope and
the disk, and in the second case, without
the disk, we adjust only the first 4 parameters for the envelope. 
The fitting is
done independently in each case, and the 4 envelope
parameters are therefore different in the 2 cases.  
The procedure for fitting the data is the same as described in \citet{Keto2004}.
We use a fast simulated annealing algorithm to
adjust the model parameters to minimize
the summed squared
difference ($\chi ^2$) between the data and the model spectra.

The model spectra 
for each particular set of model parameters
are generated by our radiative transfer 
code MOLLIE \citep{Keto1990,Keto2004}. 
We assume LTE conditions for the NH$_3$ and CH$_3$CN lines.
The LTE approximation is appropriate for NH$_3$
because in the absence of strong infrared radiation
the level populations are expected to be mostly in the "metastable" states
which are the lowest J-state of each K-ladder. Since radiative transitions
between K-ladders are forbidden, the coupling between the metastable states
is purely collisional and the population approximates a Boltzmann distribution
as in LTE \citep{HoTownes1983}.  CH$_3$CN is also a
symmetric top and transitions across the K-ladders are
similarly forbidden; however, unlike NH$_3$, 
the upper states are easily 
populated in warm gas ($> 100$ K). While the justification for the LTE
approximation is not as strong for CH$_3$CN as for NH$_3$, we find
that CH$_3$CN always traces hot, very dense gas where collisional
transitions should be important. For C$^{34}$S we use the accelerated
lambda iteration algorithm (ALI) of \citet{RybickiHummer1991}
to solve for the non-LTE level populations.

In comparing the model to the data, we take into account the 
spatial averaging of the brightness by the width of the
observing beam. We  compute spectra over a grid of
positions and smooth the result by convolution with a 
Gaussian with the FWHM equivalent to the observing
beam of each observation, \S \ref{data}.

\begin{figure}
\includegraphics[width=7in]{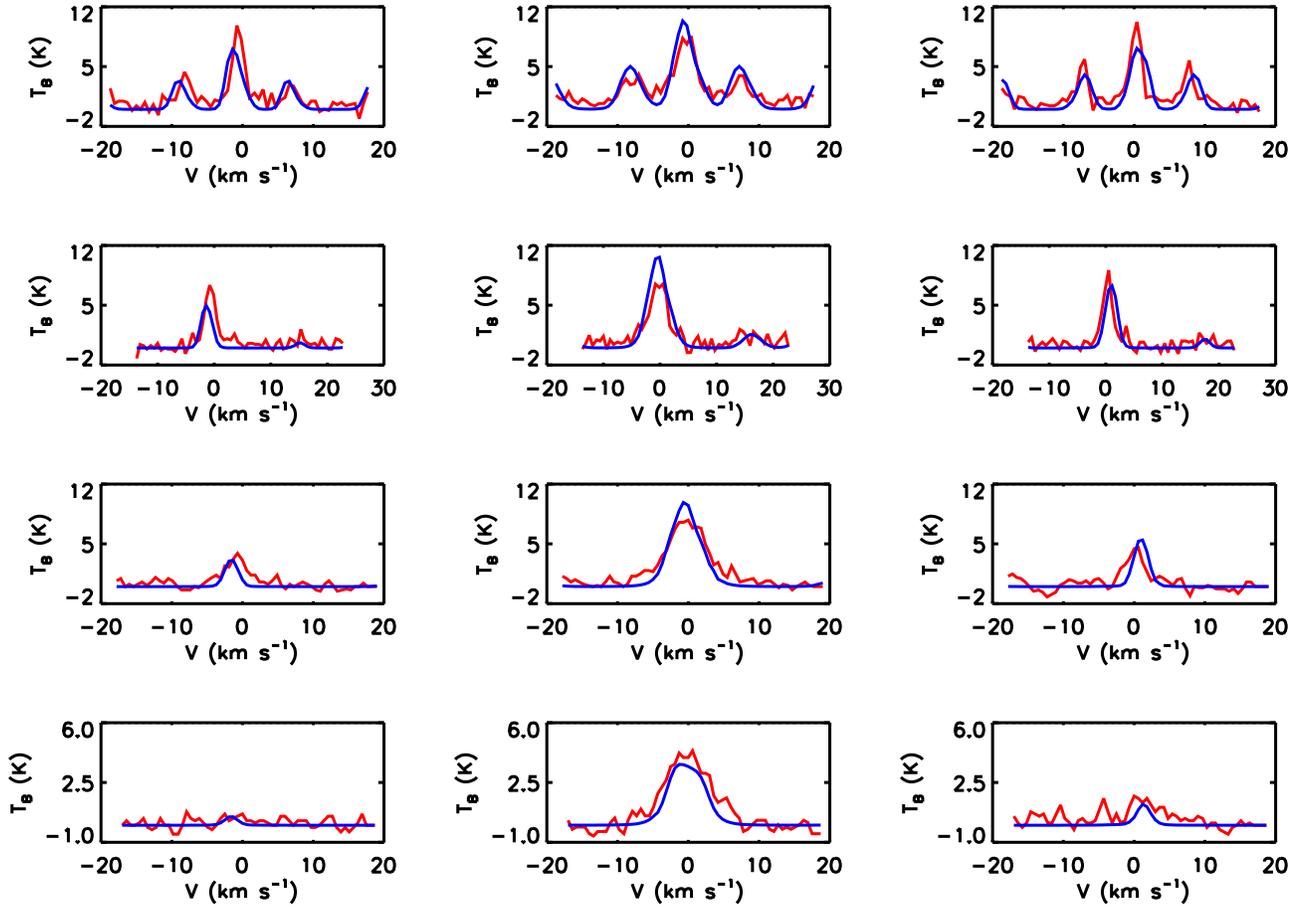}
 \caption{Data ({\it red}) and model ({\it blue}) spectra of 
 NH$_3$(1,1) ({\it top row}) NH$_3$(2,2) ({\it 2nd row})  
 NH$_3$(3,3) ({\it 3rd row}) and NH$_3$(4,4) ({\it bottom row}) at 
 3 positions across the mid-plane
 of IRAS20126  for the accretion flow including both an infalling envelope
 and a flared disk. The spectra in the middle column are toward the center
 of the flow, and the the columns to the left and right show spectra
 on either side of the center, 6200 AU to the left, and 5700 au to the right.
The 3 locations are shown in figure \ref{fig:maps}.
}
 \label{fig:nh3disk}
\end{figure}

\begin{figure}
\includegraphics[width=7in]{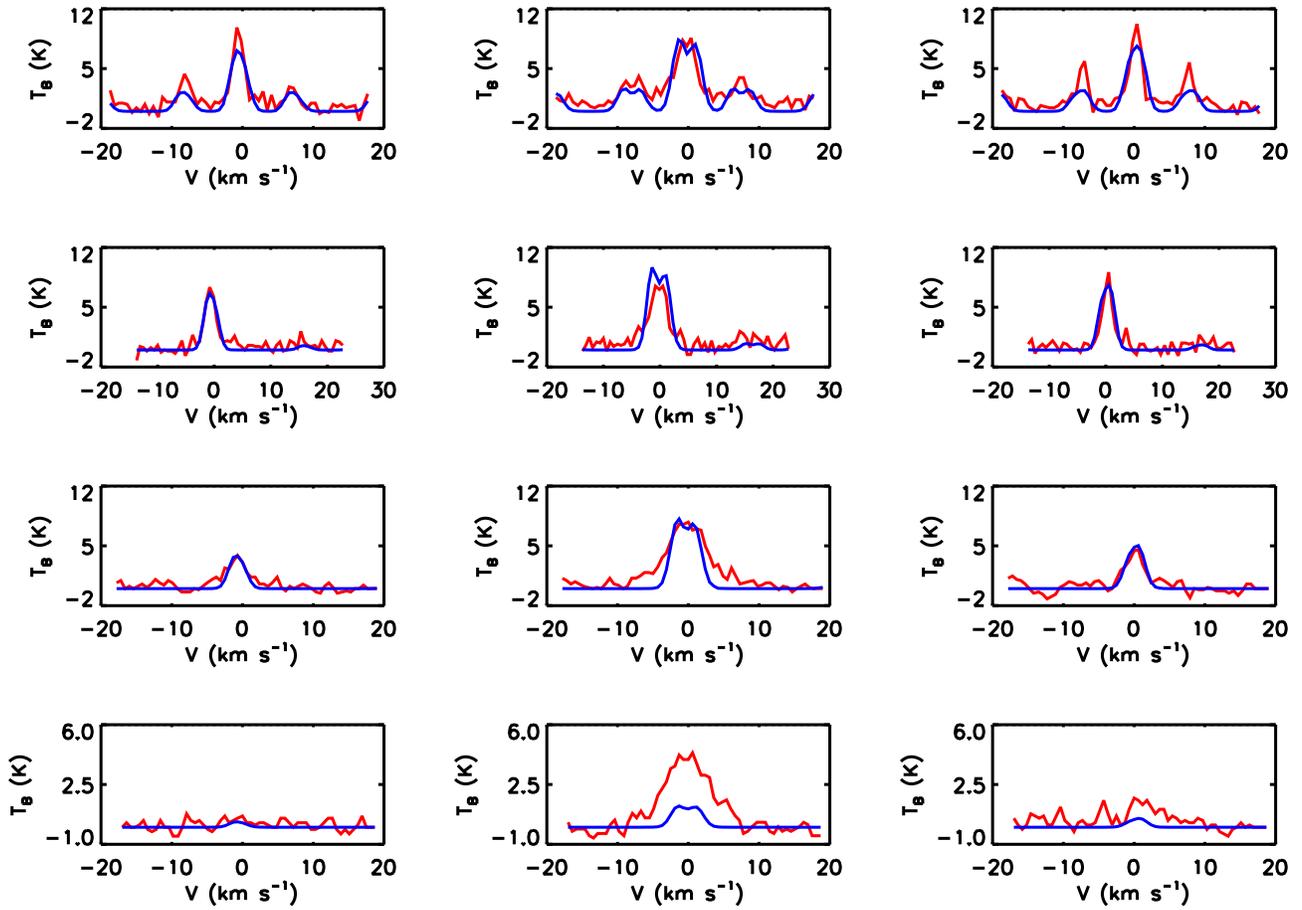}
 \caption{Spectra in the same format as figure \ref{fig:nh3disk} for the 
 accretion flow with an envelope only without a flared disk.}
 \label{fig:nh3nodisk}
\end{figure}

\begin{figure}
\vspace{0.5truein}
\centering
\mbox{\subfigure{
\includegraphics[width=3in,angle=90]{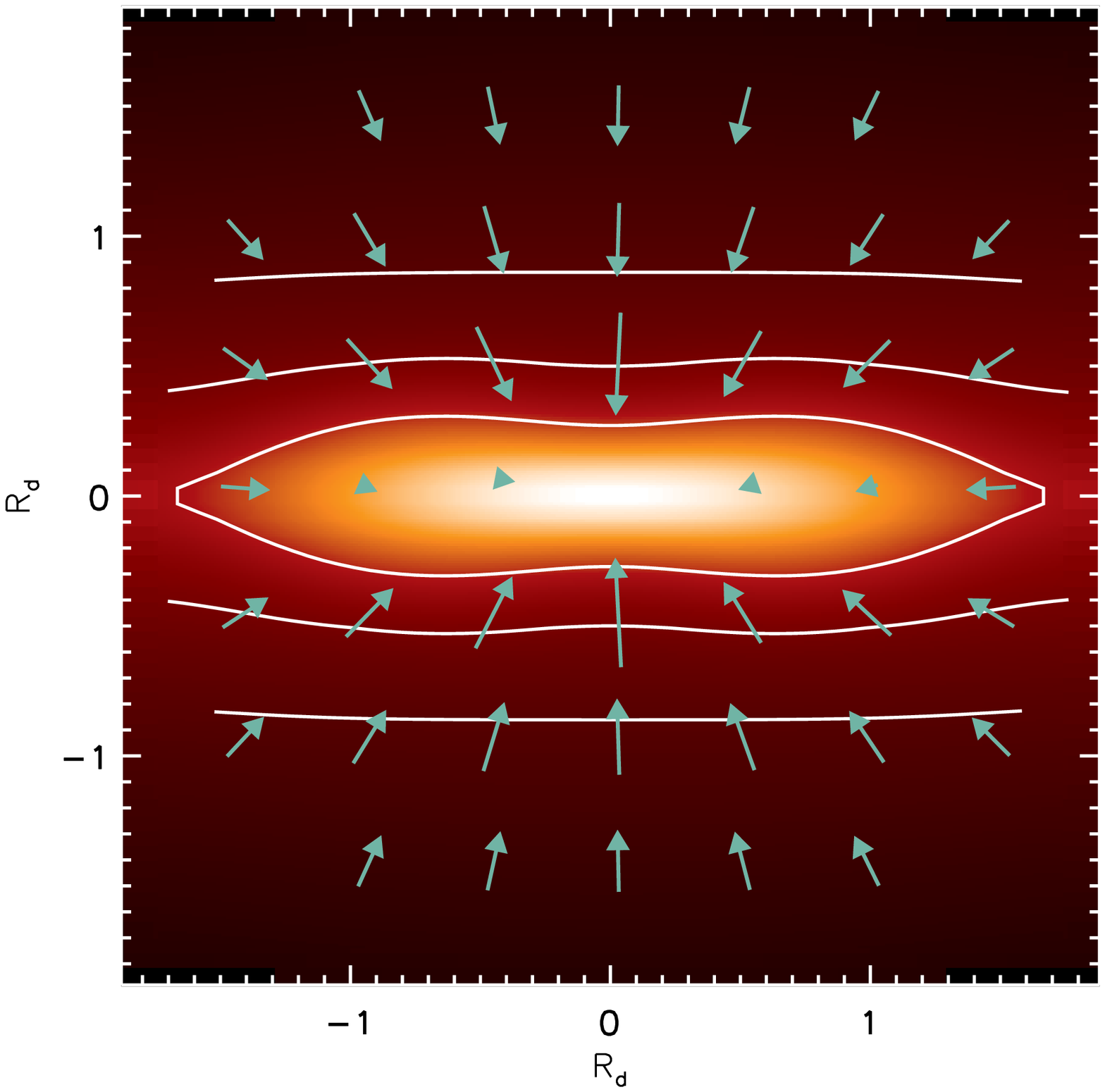}}
\subfigure{
\includegraphics[width=3in,angle=90]{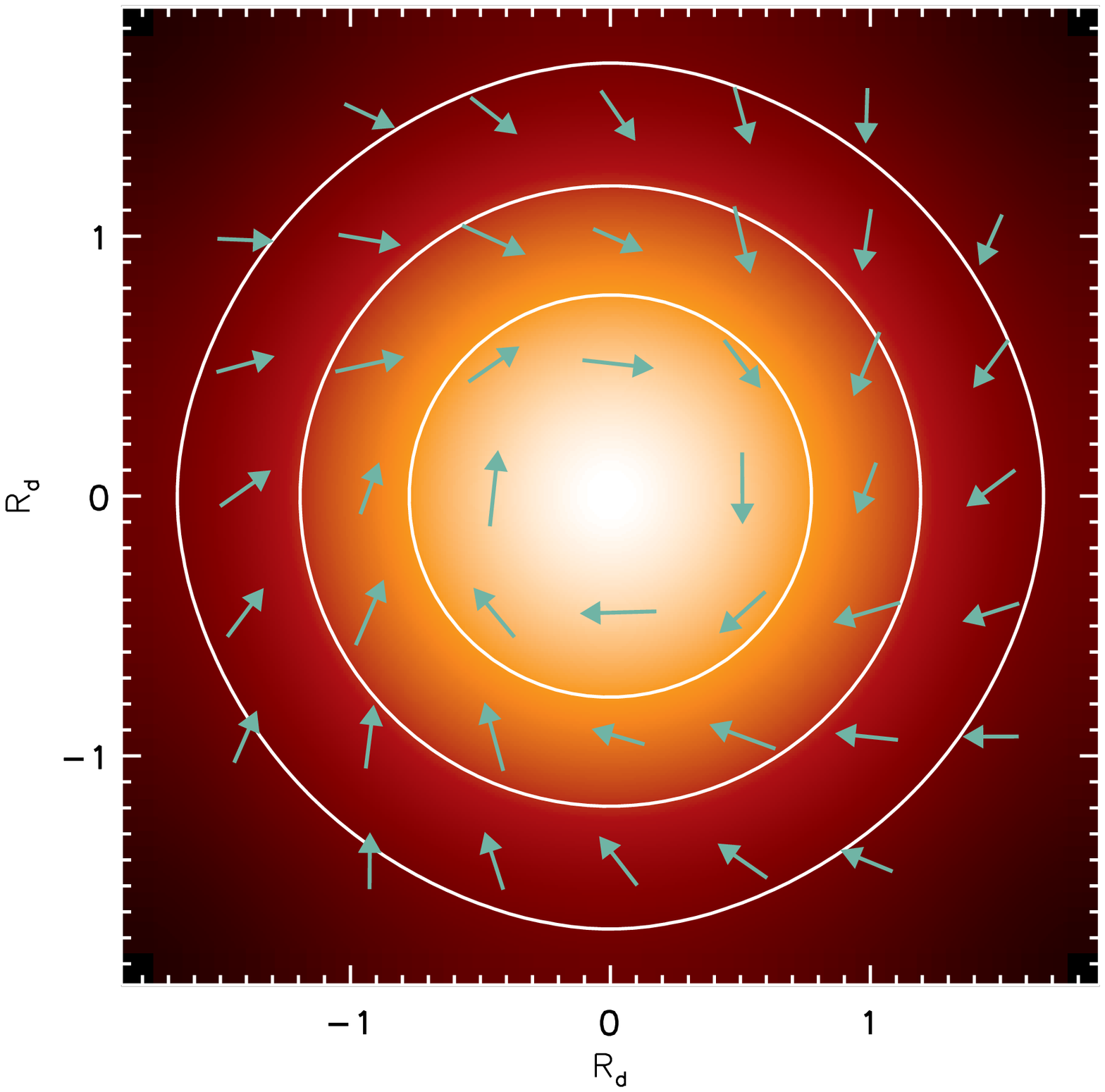}}}
 \caption{Density ({\it color}) and velocities ({\it arrows})
for the model with both an envelope and flared disk.
{\it Left:} The XZ plane (perpendicular to
 the rotation) and {\it right:} the XY plane (the mid-plane of the
 disk and rotationally-flattened envelope).
In the left panel, the color scale is  logarithmic between $9.2\times 10^4$
 and $3.7\times 10^6$ cm$^{-3}$.
 Contours ({\it white}) indicate the density at 0.006, 0.01, and 0.03, of the peak
 density. 
The longest arrow corresponds to 2.5 kms$^{-1}$
 In the right panel the color scale is 
 logarithmic between $1.1\times 10^5$
 and $8.1\times 10^7$ cm$^{-3}$.
 Contours ({\it white}) show the density at 0.03, 0.10, and 0.3 of the peak
 density. 
The linear scale in both panels is $R_d = 6900$ AU. 
}
 \label{fig:mapXZ}
\end{figure}

We simultaneously fit 12 NH$_3$ spectra, 4 transitions at
3 locations. The 3 locations are marked on figure \ref{fig:maps}.
We ran twenty thousand trial models for each of the two cases
with and without the disk.
The spectra of the 2 best-fit models (with and without the disk) 
are shown in figures \ref{fig:nh3disk}
and \ref{fig:nh3nodisk}.  Parameters for both cases are listed in table 1. 
Figure \ref{fig:TDplot} shows
the temperature and density in the mid-plane
of our best fitting models.
Figure \ref{fig:mapXZ} shows the
density and velocity of the disk-envelope model 
on planes parallel and perpendicular
to the rotation.

We also have CH$_3$CN and C$^{34}$S data from the observations
of \citet{Cesaroni1999}. The way the radiative transfer
simulation program operates, we cannot use our automated
search algorithm on more than one molecule at a time.
The program is recompiled for each molecule. We also have
not implemented the automated search for CH$_3$CN. We could
fit the model to the C$^{34}$S data, but this line does
not have the hyperfine structure of NH$_3$ that is so useful
in constraining the optical depth and temperature. Therefore
we opted for a different strategy. 
We do not use the CH$_3$CN or the C$^{34}$S data in the fitting.
We use the data on these other lines as a check on the 
model derived from the NH$_3$ data. We
simulate the CH$_3$CN and C$^{34}$S emission using the same 
two models (with and without
the disk) previously derived from
the NH$_3$ observations for comparison with the predicted spectra
against the data. The models derived from the NH$_3$ fitting
fix the temperature, densities, and velocities, but we
still need to assume abundances for CH$_3$CN and C$^{34}$S.
For CH$_3$CN
we assume an abundance 
of $6\times 10^{-8}$. 
The assumed abundance of C$^{34}$S 
is $9.0\times 10^{-11}$,
chosen to match the brightness of
the (2-1) line.

\subsection{Comparison of the model with the observed spectra}

The comparison of figures  \ref{fig:nh3disk}
and \ref{fig:nh3nodisk} shows that the higher temperature
and density disk and the lower temperature and density envelope 
are both required to fit both the high-excitation and low-excitation NH$_3$ 
spectra.  
Without the disk,
there is not enough warm, dense gas  to reproduce the (4,4) line
brightness.
Without the large-scale envelope, there is not enough cold gas
to fit the observed ratios of the NH$_3$(1,1) hyperfine lines. 

Further evidence for the disk component comes from the CH$_3$CN spectrum.
The results again show that the warm, high-density,
disk component is required to get enough brightness in lower frequency
(higher velocity) K transitions. Based on the observed line width, 
the CH$_3$CN(12-11) emission comes
from gas very close to the star. The observed CH$_3$CN(12-11) lines that are 
not blended with
other molecular lines, have line widths of 10.0 kms$^{-1}$.  
The 2 lines that appear much broader in the data contain contaminating
emission from transitions of CH$^{13}$CN(12-11)
and HNCO \citep{Cesaroni1999}.
Rotational velocities $>10$ kms$^{-1}$ are only found at 
radii closer than 600 AU ($<  G M /v^2$) around a 10 M$_\odot$ star.
Figure \ref{fig:TDplot} shows that the gas in the disk
has a temperature of greater than 150 K within this radius.  Thus the
CH$_3$CN emission derives mostly from warm gas that is very
close to the star and is thus a good molecular tracer of the
inner disk.

The CH$_3$CN(13-12) spectra is too noisy to help discriminate between
the two cases. The peak signal-to-noise ratio is about 6 after
smoothing by every other channel.
Nonetheless, the models are at least consistent
with these observations.

\begin{figure}
\vskip -0.25in
\centering
\mbox{\subfigure{
\includegraphics[width=3in]{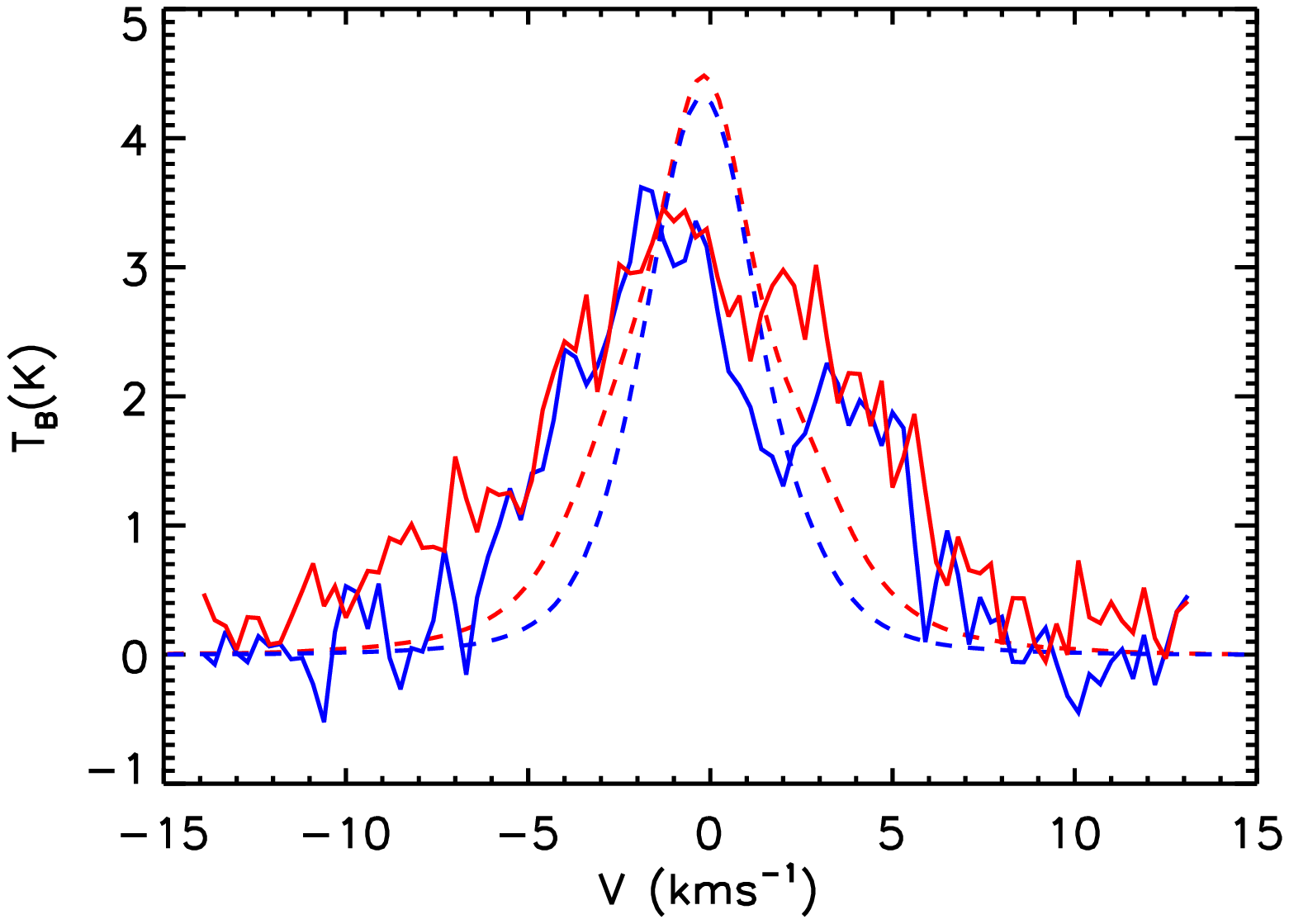}}
\subfigure{
\includegraphics[width=3in]{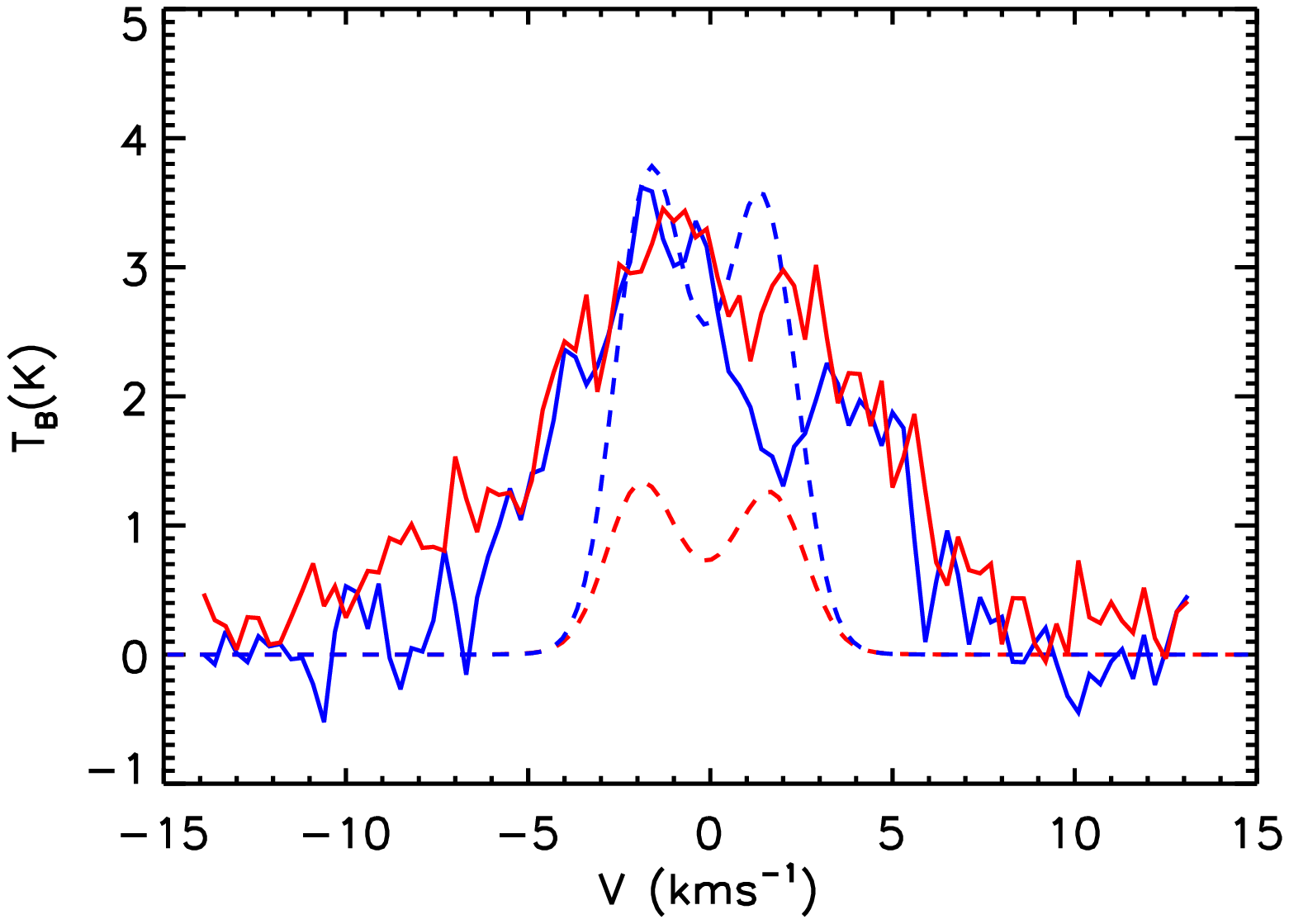}}
}
 \caption{C$^{34}$S(2-1) ({\it blue}) and C$^{34}$S(5-4)({\it red}) spectra of the data
 ({\it solid lines}) and the model ({\it dashed lines}).
 {\it Left:} Model with both the accretion envelope and flared disk.
{\it Right:} Model with only an accretion envelope and without a flared disk. 
The brightness of the C$^{34}$S(5-4) spectra, both observed and modeled,
 have been divided by 6. The comparison shows that the model with the disk
produces the correct brightness ratio. Without the disk, the C$^{34}$S(5-4)
line is not bright enough and both lines have the wrong profile shape.
The zero velocity refers to the model.
The observed spectra have been shifted to match.
} 
 \label{fig:cs_spectra}
\end{figure}

The brightness ratio of the C$^{34}$S (5-4) and (2-1) lines
also suggests the presence of a warm, dense disk
(figure \ref{fig:cs_spectra}). The model
with the disk reproduces the observed brightness ratio, 6:1 
for the 2 lines, although line width is less than observed.
The model without the disk is not able to
generate sufficient brightness in the higher excitation 
line, and
the shapes of the line profiles do not match the data.
In particular, the strong splitting that is seen 
in the model profiles without the disk
is not seen in the data. 
This difference
is also seen in the NH$_3$ and CH$_3$CN spectra of the
2 models, although not as prominently.
We usually associate split line profiles with spatially
unresolved observations of Keplerian
disks \citep{BeckwithSargent1993} whereas here the
disk produces a triangular profile. What happens here
is that the disk component is spatially resolved and has a 
very high density.
Thus within the beam through the center
of the model there is a lot of high density gas from the outer part 
of the disk
with very low velocity projected 
along the line-of-sight.  
This gas 
disk creates the peak in the spectrum around zero velocity.
In the envelope-only model without the high density disk, 
the model is optically thin in C$^{34}$S. 
In this case, we get the usual split line profiles from
the rotation and infall in the envelope.

\subsection{Goodness of fit}
How well do the data constrain the model parameters? Some appreciation
can be gained by plotting $\chi^2$ obtained from all the trial models
versus each model parameter. 
Each panel shows the fits obtained (ordinate) for each value
of a single parameter (abscissa) as all the other parameters
are varied. The lowest value of $\chi^2$ (ordinate) at each
parameter value (abscissa) is the best fit that can be
obtained for that parameter value, for any combination of all the other 
parameters.
The formal error of a model parameter is proportional
to the second derivative of $\chi^2$ with respect to the model parameter.
Thus the curvature or width of the 
lower boundary of the collection of points in each figure 
is a qualitative measure of the sensitivity
of the model to the parameter. 

\section{Discussion}

\subsection{The density}
The gas density is constrained by the optical depths of the NH$_3$
lines, which can be determined from 
the brightness ratios of the hyperfine lines. 
The density, $\rho_0$, at $R_d$ is  $7.9 \times 10^4$ (table 1) 
assuming an NH$_3$ abundance of $10^{-7}$.
With this abundance, the mass of the disk is 2.5 M$_\odot$,
the mass of the envelope is 12.6 M$_\odot$, and
the accretion rate (equation \ref{eq:accretion_rate}), 
asssumed to be the same in the
envelope and the disk, 
is $\dot{M} = 7.6\times 10^{-5}$ M$_\odot$ yr$^{-1}$.  
The total mass in the accretion flow is 
mass of the disk and envelope together which is 15.1 M$_\odot$.
This mass estimate
is the total mass within the model boundary of 26000 AU. 
\citet{Cesaroni1999} estimates
that there is between 0.6 and 8.0 M$_\odot$ within a radius of 5000 AU.

The mass estimate derived from molecular line observations
is subject to a large uncertainty.
The radiative transfer modeling determines the column density of NH$_3$
rather than H$_2$.
Therefore, the gas density, the masses of the disk
and envelope, and the accretion rate depend 
inversely on the assumed abundance of NH$_3$ which is not
well known. Estimates from models and observations of 
similar clouds range from 
$10^{-9}$  to $10^{-6}$ 
\citep{HerbstKlemperer1973,Keto1990,Estalella1993,
Caproni2000,Galvan2009}. 
Thus the abundance of NH$_3$ may be uncertain
by more than an order of magnitude, yet
a factor of 2 in 
the masses of the disk and envelope is significant in an
interpretation of the accretion dynamics. 
Furthermore, the abundance of NH$_3$ could be different
in the disk and envelope. Estimates of the NH$_3$ abundance are generally
lowest in colder clouds and highest in warm gas around
massive stars. Some NH$_3$ may be frozen onto
dust grains in colder gas and sublimated into the gas phase as the
temperature rises. Therefore, it is possible that the NH$_3$ abundance is
higher in the warm disk than the cold 
envelope. If so, the mass of the envelope could be
higher than 12.6 M$_\odot$, assuming a lower abundance,  
without necessarily 
implying a higher mass for the disk.

\subsection{The disk density factor}
\begin{figure}
\vspace{0.5truein}
\includegraphics[width=3in]{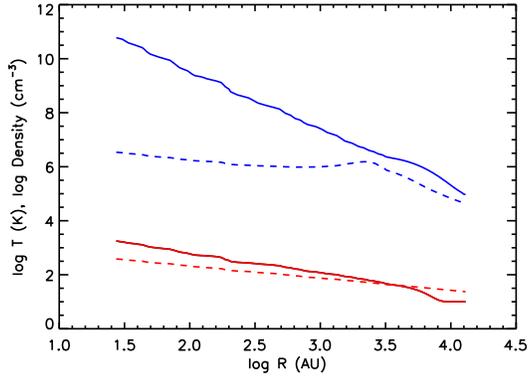}
 \caption{The temperature ({\it red}) and density ({\it blue})
 for the model with both an envelope and flared disk ({\it solid line})
 and for the model without a disk ({\it dashed line}). 
The temperatures are about the same in both
models, but the disk adds considerable density.
The increase in density in the envelope-only model (dashed)
at 3300 AU is the smoothed density singularity at $R_d$ in the
Ulrich (1976) model.
 }
 \label{fig:TDplot}
\end{figure}

In the model with the flared disk, the density in the mid-plane of
the disk 
at the radius of the disk boundary $R_d$ 
is 5 times
more dense than the smoothed density of the envelope at the same point.
Considering that the gas density changes by 6 orders of
magnitude across the model, there
is little discontinuity 
between the envelope and disk densities (figure \ref{fig:TDplot}). 
In the model without the disk the smoothed density
is approximately constant from $R_d$ to the origin
(figure \ref{fig:TDplot}).

\subsection{The stellar mass} 
The stellar mass, 10.7  M$_\odot$,
is constrained primarily by the disk velocities
(equation 17 
) required
to match the observed linewidths (about 10 kms$^{-1}$)
and the gas temperature that determines the brightness
ratios of the low and high excitation lines.
The gas velocities and therefore
line widths due to unresolved motions within the beam
depend on the stellar mass because the velocities
are proportional to the square root
of the mass. The gas temperature depends on
the mass of the star because the stellar temperature is
a strong function of the stellar mass and because
the accretion rate depends on the stellar mass.
In the disk, heating by both the star and by accretion 
(dissipation) are important
(equations \ref{eq:Tenv}, \ref{eq:Tdisk} and \ref{eq:Tavg}   ).
In the best-fit model,
the disk temperature
would be about 1/3 lower
without the
passive heating from the star, 
The temperature of the
infalling envelope is determined entirely by the
stellar temperature (equation \ref{eq:Tenv}).

\subsection{The angular momentum of the envelope}

The angular momentum of the envelope is determined
from the widths of the spectral lines and from
the VLSR of the NH$_3$ spectra at the locations
left and right of the center.
If there were too much rotation then both of the off-center spectra would have
an incorrect V$_{\rm LSR}$;  too little, and all the NH$_3$ line widths
would be too narrow.
The disk radius, $R_d = 6900$,  is set by the 
angular momentum in the envelope, $\Gamma$, and the
stellar mass (equation \ref{eq:RD}).
The velocity at $R_d$, $v_k = 1.2 $ kms$^{-1}$, increases inward as $R^{-1/2}$.
We assume a microturbulent broadening of 1 kms$^{-1}$, added
in quadrature to the thermal broadening. Since the observed line widths
are about 10 kms$^{-1}$ most of the width
of the spectral lines comes from rotation and infall in the model.

The model radius of 6900 AU seems large, and we would regard this
as an upper limit. First, the Ulrich flow conserves angular
momentum whereas real accretion flows probably involve some
braking of the spin-up. If the spin-up were slower, then the
disk would be found at a smaller radius. Second, the high
density of the disk is helpful in providing optical depth
to strengthen the NH$_3$(1,1) outer hyperfines. The Ulrich flow
has a density singularity in the mid-plane which we have 
handled as desribed in \S \ref{singularities} by a combination
of gridding and smoothing. Maybe our mid-plane density in the
envelope is too low, and to compensate, the thin disk is bigger.

\subsection{The exponent of the temperature power law}
The modelling suggests that the temperature of the envelope
falls off very quickly away from the star, as
$r^{-1}$. This implies that the dilution of the 
radiation is faster than spherical, $r^{-2}$.
The rapid dilution is consistent with the rotationally
flattened geometry. At each radius, radiation is absorbed
and re-emitted by dust in the flow. 
In a flattened flow, much of this reprocessed radiation 
escapes vertically out of the 
flow. In contrast, in a spherical flow, all the reprocessed radiation
continues to interact with the flow at larger radii.
If the gas temperature decreases outward fast
enough that most of the envelope is at the minimum gas temperature,
then a larger value of the exponent (more negative value of
the parameter $p$ in equation \ref{eq:Tenv})  would have no
further effect. 
We assume a minimum temperature of 10 K, 
a typical temperature of molecular gas,
We derive an upper limit $p < -1$.

\subsection{The disk temperature multiplier}
The disk temperature multiplier determines the
relative importance of active disk heating, owing to
accretion and dissipation (equation \ref{eq:Tdisk}), and 
passive heating from
the protostar (equation \ref{eq:Tenv}).
With active disk heating, the temperature in the
disk is about 50\% above that from passive heating 
alone.

\section{Conclusions}

This investigation shows that a standard model of a freely-falling,
rotationally-flattened accretion envelope around a rotationally supported
disk is able to reproduce the NH$_3$, CH$_3$CN, and C$^{34}$S spectral
line observations of IRAS20126. 
Both the  disk and envelope components are required to fit the
observed brightness of both the
low and high excitation lines, and to fit the observed line widths.

This disk-envelope model was developed for low-mass stars and is
quite successful in explaining many of their observable characteristics.
The success of this model in explaining the molecular line observations
of the massive star IRAS20126 suggests that at least up
to  10 M$_\odot$, the accretion processes of massive stars 
are similar to those of solar mass stars.

There are some differences. Although the mass of the disk is uncertain
owing to its dependence on the NH$_3$ abundance, the disk mass 
is a significant fraction of the mass of the star. Furthermore, the
extent of the disk is quite large, $\sim$6900 AU. This suggests that
self-gravity in the disk is dynamically important, and therefore, the disk may
be unstable to local fragmentation and the formation of companion
stars. 

The physical model in combination with the molecular line
radiative transfer
presented in this paper has a further application to other
massive star forming regions. The spatial resolution
and sensitivity of the present generation of interferometers cannot spatially
resolve accretion disks in massive protostellar objects at
multi-kpc distances. However, spectral lines sampling 
a wide range of densities and temperatures can still provide
constraints to the physical structure of the core and disk. As
shown in IRAS20126, the spectral lines formed in the
higher density and temperature regions confirm the presence of
an accretion disk.  
While future observatories such as ALMA will be able to spatially
resolve the flow on the disk scale,
before the science commissioning of ALMA, this method
is a promising technique to probe the spatially unresolved regions of
the flow and disk
using data from the current interferometers.

\section*{Acknowledgments}
The authors thank Riccardo Cesaroni and T.K. Sridharan for 
the use of their observational data.


\bsp

\begin{table*}
 \centering
 \begin{minipage}[t]{3truein}
  \caption{Model Parameters.}
  \begin{tabular}{@{}llll@{}}
  \hline
   Parameter     &Symbol &Disk            & No Disk\\
 \hline
 Env. Density at $R_d$ (cm$^{-3}$)		&$\rho_0$	&$7.9\times 10^4$  	&$7.9\times 10^5$  	\\
 Temperature power law exp.			&$p$		&$<-1$	&0.4		\\
 Angular Momentum	(AU kms$^{-1}$)	&$\Gamma$	&8100 	&3500	\\
 Stellar Mass (M$_\odot$)				&M$_\ast$	&10.7	&7.3		\\
 Disk Density Ratio					&$A_\rho$	&5.1		&...		\\
 Disk Temperature factor				&$B_T$		&15.0  	&...		\\
Centrifugal radius (AU)					&$R_d$	&6900	&1900	\\
Velocity at $R_d$ (kms$^{-1}$)			&$v_k$		&1.2		&1.8		\\
Total mass$^1$ within 0.128 pc ($M_\odot$)&			&12.6	&10.1	\\
Disk mass ($M_\odot$)				&			&2.5		&...		\\
Envelope accretion rate ($M_\odot$ yr$^{-1}$)	&		&$7.6\times 10^{-5}$	&$1.0\times 10^{-4}$\\
\hline
$^1$ The mass estimates include the envelope and the disk for the case & \\
with a disk, and the envelope only for no disk. & \\
Both cases assume an NH$_3$ abundance of $1.0\times 10^{-7}$. & \\
\end{tabular}
\end{minipage}
\end{table*}

\onecolumn

\clearpage

\clearpage

\clearpage

\begin{figure}
\vskip -0.25in
\centering
\mbox{\subfigure{
\includegraphics[width=3in]{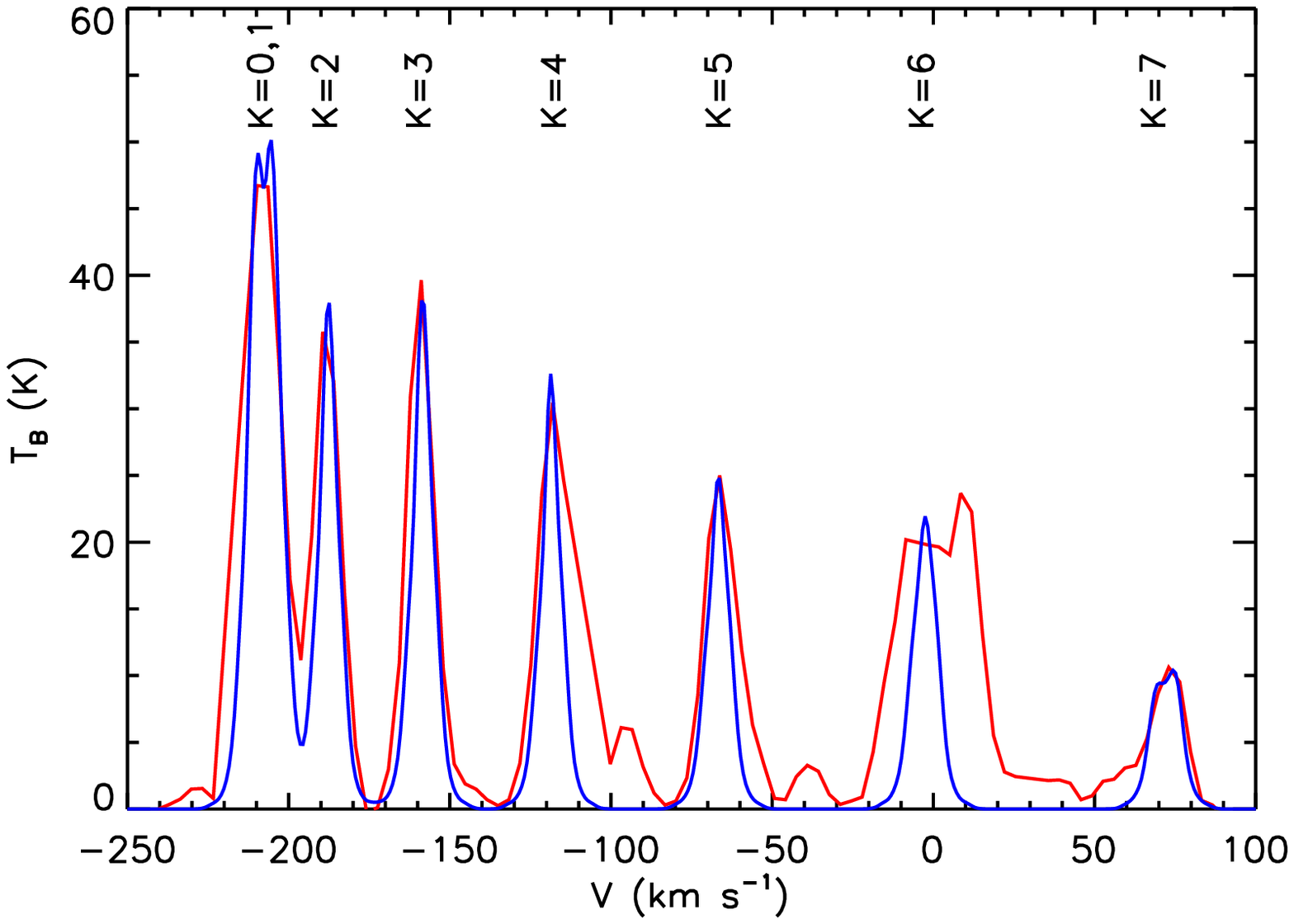}}
\subfigure{
\includegraphics[width=3in]{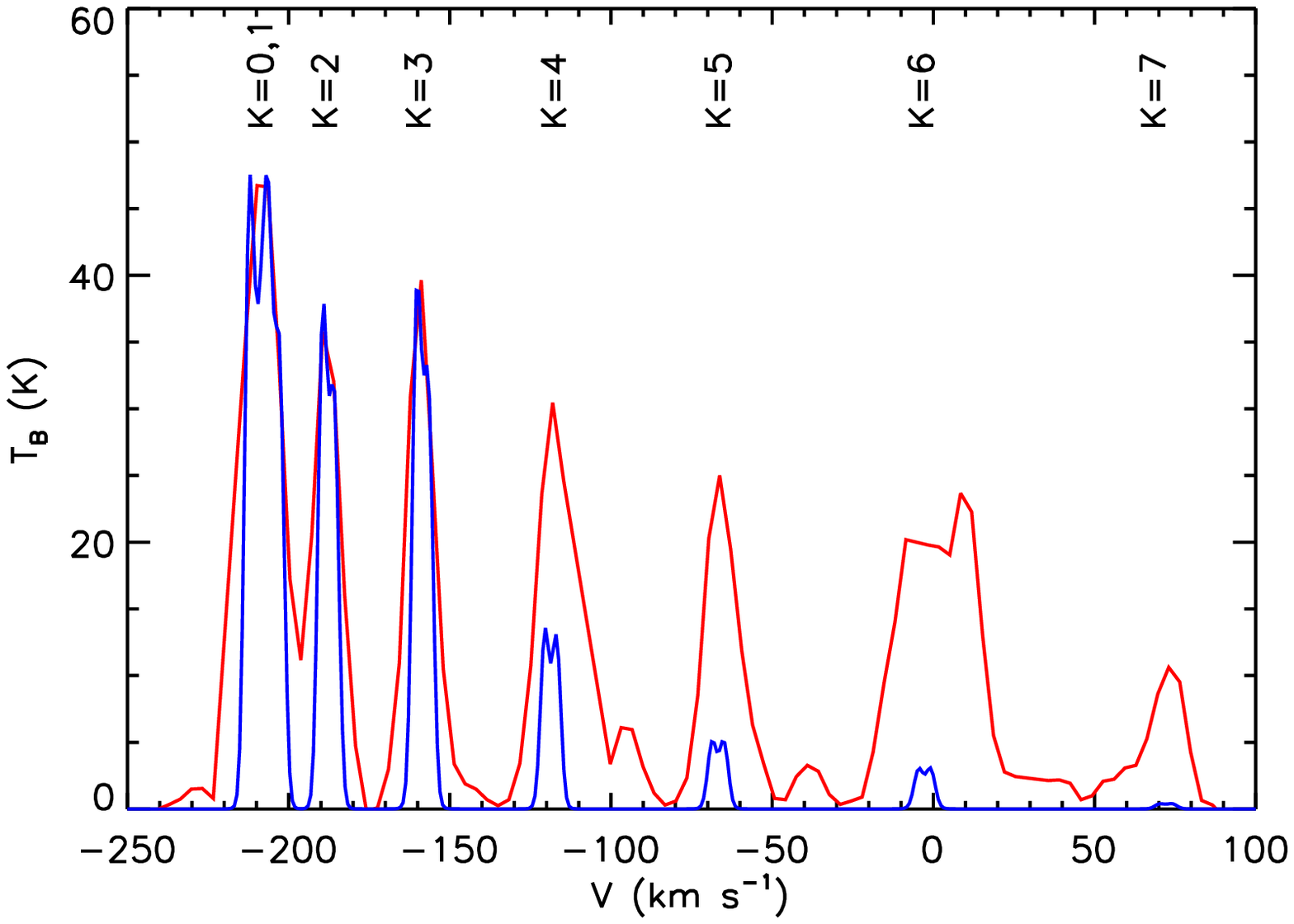}}}
 \caption{Data ({\it red}) and model ({\it blue}) spectra of CH$_3$CN(12-11).
The {\it left} panel shows the spectrum for the model
 with both an accretion envelope and flared disk. The model is the same as
 shown in figures \ref{fig:nh3disk}, \ref{fig:TDplot}, \ref{fig:mapXZ}. 
The {\it right} panel is for the model with only an accretion envelope and
without a disk (figures \ref{fig:nh3nodisk}, \ref{fig:TDplot}).
In the observed spectrum, the 
line that is anomalously wide (K=6, 2nd from right) is contaminated by
 HNCO emission. Additionally, contamination from weak CH$_3 ^{13}$CN emission
shows up to the
 red (right) of 4 of the lines. 
 The velocity of the K=6 hyperfine line has been set to zero.
The observed spectra have been shifted in velocity to match.
}
 \label{fig:ch3cndisk}
\end{figure}

\begin{figure}
\vskip -0.25in
\centering
\mbox{\subfigure{
\includegraphics[width=3in]{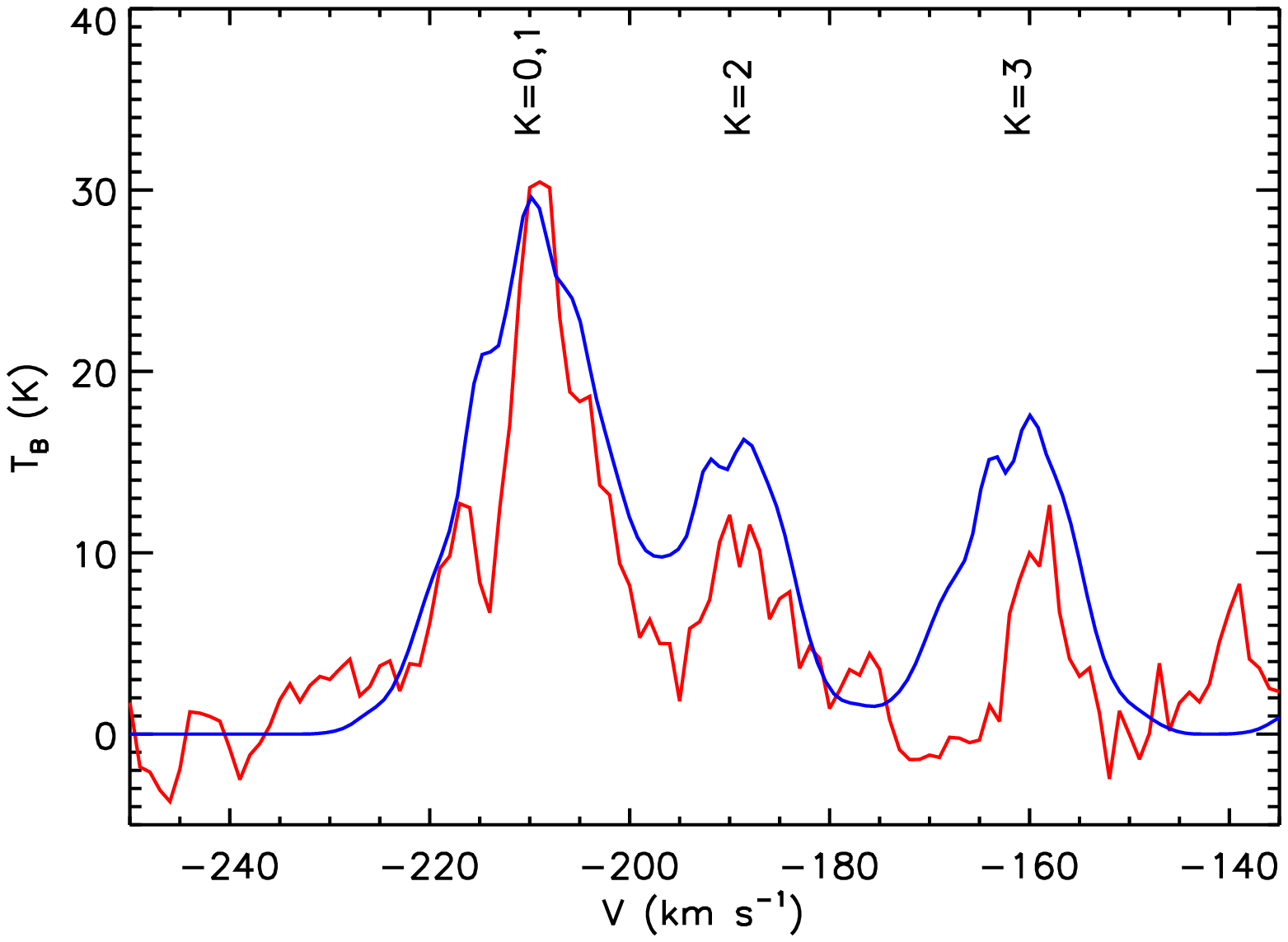}
}
\subfigure{
\includegraphics[width=3in]{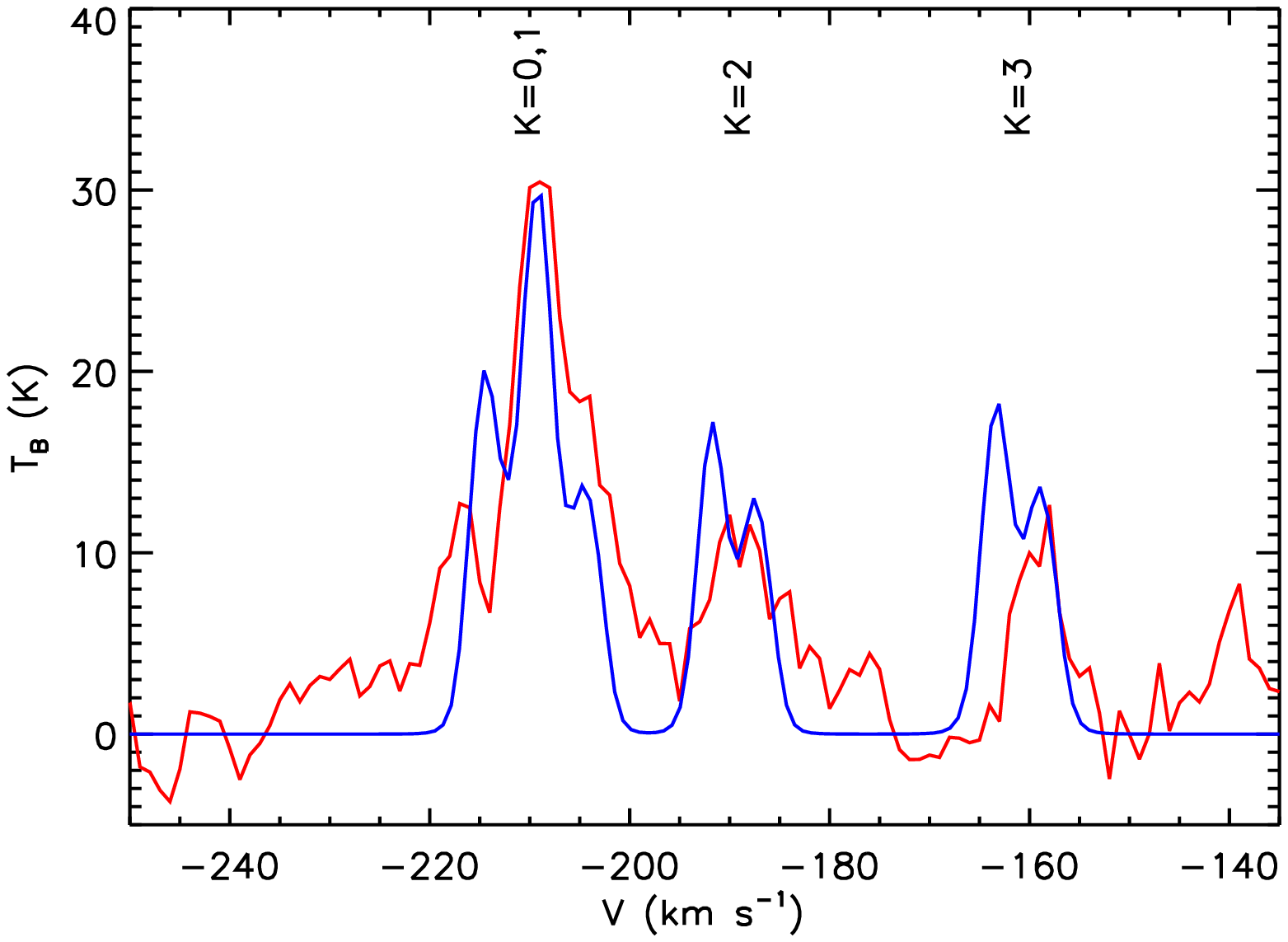}
}}
 \caption{
Spectra of CH$_3$CN(13-12) in the same format as figure \ref{fig:ch3cndisk}
except that we have only the 4 lowest K transitions. The K=0 and K=1
transitions are blended together in the bluest peak. 
The {\it left} panel shows the spectrum for the model
 with both an accretion envelope and flared disk. 
The {\it right} panel is for the model with only an accretion envelope and
without a disk (figures \ref{fig:nh3nodisk}, \ref{fig:TDplot}).
The signal-to-noise ratio of the observed spectra, about 6 at the peak,
is not high enough to discriminate between the two models. At least
the models are consistent with the data. The data are from 
T.K. Sridharan (private communication).
}
 \label{fig:ch3cnnodisk}
\end{figure}


\clearpage

\begin{figure}
\vspace{0.5in}
\includegraphics[width=7in]{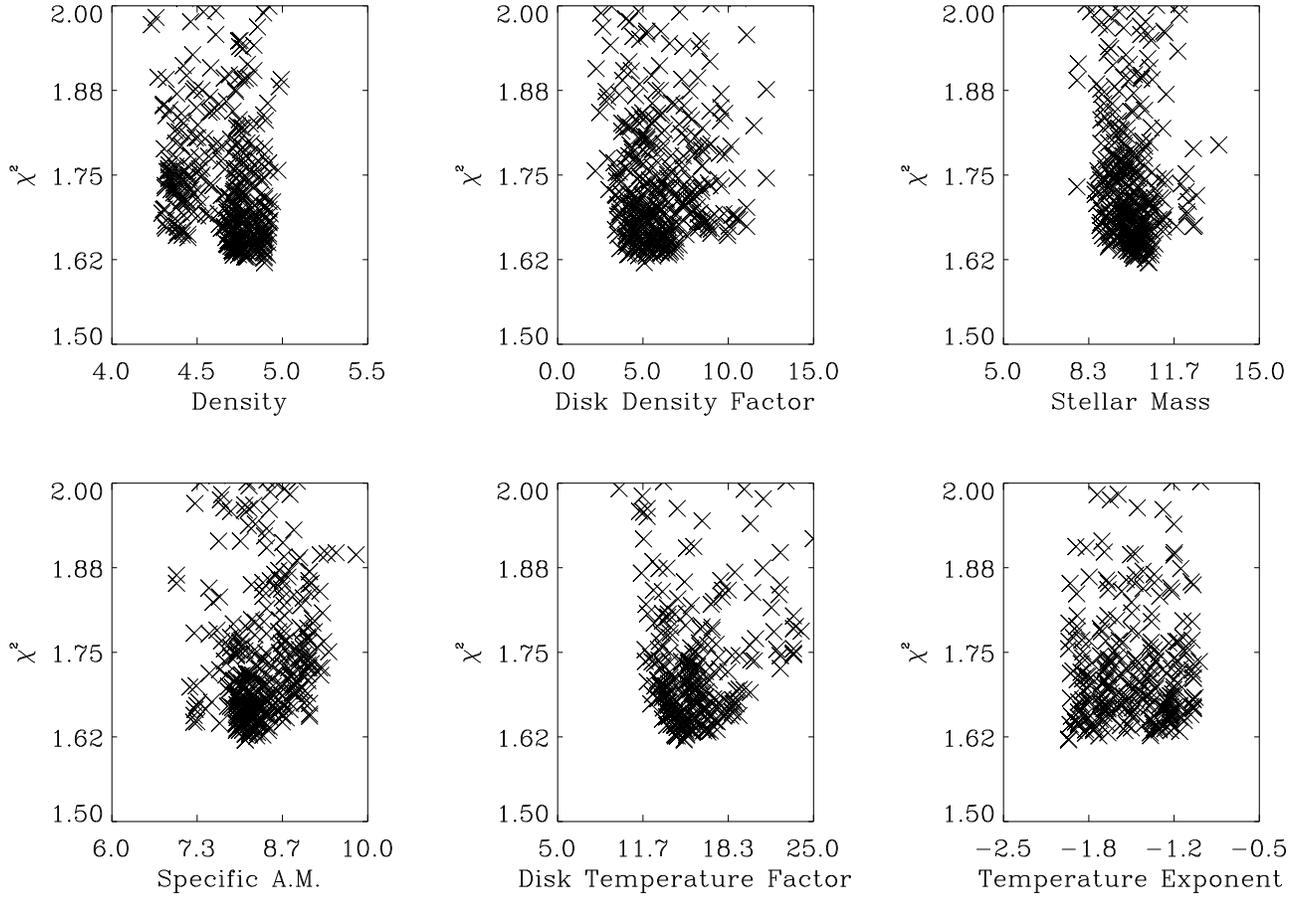}
 \caption{The value of $\chi ^2$ as a function of a single parameter.
 The plots show $\chi^2$ for the better fitting disk-envelope
 models as a function of
 a single parameter. The width of the collection of points in each
 figure is a qualitative
 measure of the sensitivity of the models to each parameter.
 Values of the parameters outside the range shown by the lower $\chi^2$
values produce very poor fits. The units of the density are the log of $cm^{-3}$,
the units of stellar mass are $M_\odot$, and the units of angular momentum
are AU $\times$ kms$^{-1}$ divided by 1000. The parameters are listed in Table 1.}
 \label{fig:parametersdisk}
\end{figure}

\clearpage

\label{lastpage}

\end{document}